# Visualization of ferroelectric domains in boracite using emission of terahertz radiation

Yuto Kinoshita[1], Noriaki Kida[1], Masato Sotome[1], Ryotaro Takeda[1], Nobuyuki Abe[1], Mitsuru Saito[2], Taka-hisa Arima[1], and Hiroshi Okamoto[1]

[1]*Department of Advanced Materials Science, The University of Tokyo, Kashiwa, Chiba 277-8561, Japan*

[2]*Institute of Multidisciplinary Research for Advanced Materials, Tohoku University, Sendai 980-8577, Japan*

Abstract

We report on the emission of terahertz radiation by irradiation of femtosecond laser pulses in non-centrosymmetric paraelectric and ferroelectric phases of $Co_3B_7O_{13}I$ boracite. The Generation of the terahertz waves in both phases is caused by optical rectification via a second-order nonlinear optical effect. In the ferroelectric phase, we successfully visualized ferroelectric domains by analyzing the polarization state of the terahertz wave radiated from the crystal. In a large area of the crystal (~500 × 500 μm$^2$), the observed polarization vector of the radiated terahertz wave was tilted from directions of spontaneous polarization, i.e., $[100]_{cub}$, $[010]_{cub}$, and $[001]_{cub}$ in cubic setting, which can be explained by the presence of a ferroelectric 90° domain wall of the $(101)_{cub}$ plane.

# 1. Introduction

Ferroelectric materials have been widely used in electronic and optical devices.[1,2)] The control of ferroelectric domains is important for achieving a high performance in those devices.[3,4)] One successful example of ferroelectric domain engineering is periodically poled LiNbO$_3$ (PPLN) nonlinear optical devices for wavelength conversion[5-7)]; periodically inverted ferroelectric domains in LiNbO$_3$ realize the quasi-phase-matching (QPM) condition in the process of second-harmonic generation (SHG), resulting in a high conversion efficiency. The possible orientation of ferroelectric polarization is determined by the symmetry of a crystal.[2,8)] However, actual ferroelectric domain patterns are often complicated, since they are strongly affected by various factors such as crystal shapes, impurities, and defects.[2,8,9)] Thus, the visualization of ferroelectric domain topology is expected to provide valuable information on the properties of ferroelectric materials and crucial for the applications.

Thus far, various methods using electron, X-ray, and optical beams have been developed to detect ferroelectric domain topology.[2)] Among them, mapping the SHG intensity[2,10,11)] is widely used for ferroelectric domain imaging over a wide range of samples. However, it is usually difficult to determine the direction of ferroelectric polarization owing to the loss of phase information of second-harmonic waves in the detection of intensity. Recently, spatial configurations of ferroelectric domains in BiFeO$_3$ thin film[12)] and croconic acid (4,5-dihydroxy-4-cyclopentene-1,2,3-trion)[13)] have been visualized using terahertz waves radiated from the crystal by irradiation of femtosecond laser pulses. The phase of the radiated terahertz waves depends on the direction of spontaneous polarization. Thus, ferroelectric domain orientations can be easily determined by mapping out the amplitude of terahertz waves. In addition, vector imaging of ferroelectric polarization can be achieved by analyzing the polarization states of the radiated terahertz waves, which is successfully demonstrated in ferroelectric 180° domains of croconic acid.[13]

To prove the versatility of the terahertz vector imaging method as a new method of detecting various types of ferroelectric domain, here, we focus on a ferroelectric oxide, $Co_3B_7O_{13}I$ boracite.[14] Boracite is one prototypical example of a ferroelectric with ferroelectric 90° domains.[15] At room temperature, $Co_3B_7O_{13}I$ has a cubic structure with the point group of $T_d$ ($\bar{4}3m$),[16,17] as illustrated in Fig. 1(a). Below 193 K, it undergoes structural phase transition from cubic to orthorhombic with the point group of $mm2$ ($C_{2v}$).[18] The schematic of the crystal structure in the orthorhombic phase is shown in Fig. 1(b). From the viewpoint of crystallography, the $[001]_{\text{orth}}$ direction of the orthorhombic phase is equivalent to the $[100]_{\text{cub}}$, $[010]_{\text{cub}}$, and $[001]_{\text{cub}}$ directions of the cubic phase.[15] In the following, we use the expression $[001]_{\text{orth}} \parallel [001]_{\text{cub}}$ from a previous work.[15] According to the structural analysis using neutron diffraction at 51 K,[18] Co atoms occupy three crystallographically nonequivalent sites in the orthorhombic phase: Co(1), Co(2), and Co(3), which are shown by red, blue, and green circles, respectively, in Fig. 1(c). Their relative positions are different from each other, but Co atoms are totally distorted along the $[001]_{\text{cub}}$ direction. On the other hand, half of I atoms (purple circles) are distorted along the $[111]_{\text{cub}}$ direction and the other half along the $[\bar{1}\bar{1}1]_{\text{cub}}$ direction.[15] As a result, $Co_3B_7O_{13}I$ shows spontaneous polarization (~2 μC/cm² at 190 K)[19] along the $[001]_{\text{cub}}$ direction. Since all the directions of the cubic crystal are equivalent, spontaneous polarization can also emerge along the $[010]_{\text{cub}}$ and $[001]_{\text{cub}}$ directions.[15] Ferroelectric domains in related materials of $Co_3B_7O_{13}I$, such as $Ni_3B_7O_{13}Br$ and $Ni_3B_7O_{13}Cl$, have been visualized using birefringence.[15]

However, the interpretation of the contrast in the birefringence image is complicated, since six polarization directions are possible in those materials. Below 37.5 K, $Co_3B_7O_{13}I$ is also known to show a linear magnetoelectric effect.[19)]

In this paper, we report on the observation of terahertz radiation in non-centrosymmetric paraelectric and ferroelectric phases of $Co_3B_7O_{13}I$ by irradiation of femtosecond laser pulses. Terahertz radiation in both phases was well described by optical rectification of the laser pulse via a second-order nonlinear optical effect. In the ferroelectric phase, we visualized ferroelectric 90° domains in a large area of the crystal (~500×500 μm$^2$) by mapping out the polarization vector of radiated terahertz waves. The observed terahertz vector image indicates the existence of a ferroelectric 90° domain wall of the $(101)_{cub}$ plane.

## 2. Second-order nonlinear optical susceptibility of $Co_3B_7O_{13}I$

In this section, we briefly explain the general mechanism of terahertz radiation in non-centrosymmetric media and describe the expression of the second-order nonlinear optical susceptibility tensor $\chi^{(2)}$ in the paraelectric and ferroelectric phases of $Co_3B_7O_{13}I$. When a non-centrosymmetric crystal is irradiated by femtosecond laser pulses with a finite bandwidth of ~10 THz, nonlinear polarization is induced by a second-order nonlinear optical process.[20)] This results in emission of terahertz radiation into free space. This is called optical rectification in a broad sense,[21)] which was recognized as the general mechanism of terahertz radiation in various non-centrosymmetric media such as ZnTe and $LiNbO_3$.[20)]

### 2. 1. Second-order nonlinear optical susceptibility in paraelectric phase

The point group in the non-centrosymmetric paraelectric phase of $Co_3B_7O_{13}I$ is $T_d$ ($\bar{4}3m$).[16,17] Within the electric-dipole approximation, the nonzero tensor components of $\chi^{(2)}$ are $\chi^{(2)}_{xyz} = \chi^{(2)}_{yzx} = \chi^{(2)}_{zxy} = \chi^{(2)}_{xzy} = \chi^{(2)}_{yxz} = \chi^{(2)}_{zyx}$.[22] Here, the $x$-, $y$-, and $z$-axes correspond to the $[100]_{cub}$, $[010]_{cub}$, and $[001]_{cub}$ directions, respectively. By using the contradicted $d$ tensors, the second-order nonlinear polarization $P^{(2)}$ is expressed as

$$P^{(2)} = \varepsilon_0 \begin{bmatrix} 0 & 0 & 0 & d_{14} & 0 & 0 \\ 0 & 0 & 0 & 0 & d_{14} & 0 \\ 0 & 0 & 0 & 0 & 0 & d_{14} \end{bmatrix} \begin{bmatrix} E_x^2 \\ E_y^2 \\ E_z^2 \\ 2E_yE_z \\ 2E_zE_x \\ 2E_xE_y \end{bmatrix}. \quad (1)$$

Here, $E_x$, $E_y$, and $E_z$ represent $x$, $y$, and $z$ components of the electric field of femtosecond laser pulses, respectively. $\varepsilon_0$ is the dielectric constant in vacuum. We used the laboratory coordinate as the $X$- and $Y$-axes, as illustrated in Fig. 2(a). $\theta$ is defined as the angle of the $[001]_{cub}$ direction of the crystal relative to the $X$-axis. In the case of the $(110)_{cub}$-oriented crystal, $P^{(2)}$ as a function of $\theta$ can be expressed as[23]

$$P^{(2)} = \begin{bmatrix} P_{110} \\ P_{1\bar{1}0} \\ P_{001} \end{bmatrix} = -\varepsilon_0 d_{14} E_0^2 \begin{bmatrix} 0 \\ \sin 2\theta \\ \sin^2\theta \end{bmatrix}, \quad (2)$$

where $E_0$ represents the amplitude of the electric field of femtosecond laser pulses.

### 2. 2.Second-order nonlinear optical susceptibility in ferroelectric phase

The crystal symmetry of the ferroelectric phase of $Co_3B_7O_{13}I$ belongs to the point group of $C_{2v}$ ($mm2$).[18] Nonzero tensor components of $\chi^{(2)}$ in the ferroelectric phase are $\chi^{(2)}_{z'x'x'}, \chi^{(2)}_{z'y'y'}, \chi^{(2)}_{z'z'z'}, \chi^{(2)}_{y'y'z'} = \chi^{(2)}_{y'z'y'}, \chi^{(2)}_{x'z'x'} = \chi^{(2)}_{x'x'z'}$.[22] Here, the $x'$-, $y'$-, and $z'$-axes represent the $[100]_{orth}$, $[010]_{orth}$, and $[001]_{orth}$ directions, respectively. Note that the $x'$-, $y'$-, and $z'$-axes also correspond to the $[110]_{cub}$, $[\bar{1}10]_{cub}$, and $[001]_{cub}$ directions of the crystal, respectively. Then, $P^{(2)}$ is expressed as

$$P^{(2)} = \varepsilon_0 \begin{bmatrix} 0 & 0 & 0 & 0 & d_{15} & 0 \\ 0 & 0 & 0 & d_{24} & 0 & 0 \\ d_{31} & d_{32} & d_{33} & 0 & 0 & 0 \end{bmatrix} \begin{bmatrix} E_{x'}^2 \\ E_{y'}^2 \\ E_{z'}^2 \\ 2E_{y'}E_{z'} \\ 2E_{z'}E_{x'} \\ 2E_{x'}E_{y'} \end{bmatrix}. \quad (3)$$

$E_{x'}$, $E_{y'}$, and $E_{z'}$ represent $x'$, $y'$, and $z'$ components of the electric field of femtosecond laser pulses, respectively. The angle $\phi$ was defined by the angle of $E_0$ relative to the $X$-axis. In the case of the $(110)_{cub}$-oriented crystal, $P^{(2)}$ as a function of $\phi$ can be expressed as

$$P^{(2)} = \begin{bmatrix} P_{110} \\ P_{1\bar{1}0} \\ P_{001} \end{bmatrix} = \varepsilon_0 E_0^2 \begin{bmatrix} 0 \\ \sqrt{2} d'_{24} \sin\phi \cos\phi \\ d'_{32} \sin^2\phi + d'_{33} \cos^2\phi \end{bmatrix}. \quad (4)$$

When the $[001]_{orth}$ direction is chosen as the $[100]_{cub}$ or $[010]_{cub}$ direction, $P^{(2)}$ is expressed as

$$P^{(2)} = \begin{bmatrix} P_{110} \\ P_{1\bar{1}0} \\ P_{001} \end{bmatrix} =$$

$$\pm \frac{\varepsilon_0 E_0^2}{4\sqrt{2}} \begin{bmatrix} \sqrt{2}(d''_{15} - d''_{24} - d''_{31} + d''_{32}) \sin 2\phi + 2(d''_{15} + d''_{24} - d''_{33})\sin^2\phi \\ \qquad\qquad\qquad\qquad\qquad\qquad\qquad\qquad -(d''_{31} + d''_{32})(1 + \cos^2\phi) \\ \sqrt{2}(d''_{15} - d''_{24} + d''_{31} - d''_{32}) \sin 2\phi + 2(d''_{15} + d''_{24} + d''_{33})\sin^2\phi \\ \qquad\qquad\qquad\qquad\qquad\qquad\qquad\qquad +(d''_{31} + d''_{32})(1 + \cos^2\phi) \\ 2\sqrt{2}(d''_{15} - d''_{24})\sin^2\phi + 2(d''_{15} + d''_{24}) \sin 2\phi \end{bmatrix} \quad (5).$$

The plus sign denotes the case for $[001]_{orth} \parallel [100]_{cub}$, whereas the negative sign, for $[001]_{orth} \parallel [010]_{cub}$.

## 3. Experimental methods

Single crystals of $Co_3B_7O_{13}I$ were grown by the flux method and characterized by X-ray diffraction measurements. The crystal orientation was checked using a four-circle X-ray diffractometer. We used a 100-μm-thick $(110)_{cub}$-oriented single crystal.

For optical measurements, we polished the surface of the sample using Al powder. To prevent sample damage by the polishing procedure, the sample was implanted into a 500-μm-thick polymer. In the terahertz radiation experiments, a mode-locked Ti:Sapphire laser (central wavelength of 800 nm, repetition rate of 80 MHz, and pulse width of 100 fs) was used to irradiate the sample in normal incidence with the spot diameter of 25 μm. The laser power was fixed to 40 mW (~100 μJ/cm$^2$ per pulse). Emitted terahertz waves were detected in transmission geometry by the standard photoconducting sampling technique with a low-temperature-grown GaAs (LT-GaAs) detector [Fig. 2(a)].[20] We defined the laboratory coordinate as $X$-axis and $Y$-axis. We detected only the $X$-axis component of terahertz waves using a wire grid polarizer. Experimental details are described in Sects. 4. 1. and 4. 2.

In the terahertz radiation imaging experiments, we obtained images by measuring the amplitude of terahertz waves at each position by the raster scan method. The sample holder was attached to a two-dimensional moving stage and moved along the $X$- and $Y$-axes. The pump pulses were focused by a lens (focal length $f$ = 50 mm). The spatial resolution was evaluated to be ~25 μm. All images were taken at 50 K. The principle and experimental procedure of terahertz vector mapping are detailed in Sect. 4. 3.

## 4. Results and discussion
## 4. 1. Terahertz radiation in non-centrosymmetric paraelectric phase

Figure 2(a) shows the experiment setup of terahertz radiation for the paraelectric phase. This experiment was performed at room temperature. $E_0$ of femtosecond laser pulses was set parallel to the $X$ axis. We rotated the sample by the angle $\theta$; $\theta = 0°$ corresponds to $E_0 \parallel [001]_{\text{cub}}$, whereas $\theta = 90°$ corresponds to $E_0 \parallel [1\bar{1}0]_{\text{cub}}$. We measured the absorption coefficient $\alpha$ at 800 nm (~114 cm$^{-1}$) and estimated the penetration depth (~88 µm) of femtosecond laser pulses, which is comparable to the sample thickness (100 µm). By the irradiation of femtosecond laser pulses, we found terahertz radiation from Co$_3$B$_7$O$_{13}$I. Figure 2(c) shows the typical waveform of the radiated electromagnetic wave with $\theta = 55°$. The radiated wave mainly consists of a single-cycle terahertz pulse with a pulse width of ~1.2 ps. On the other hand, no electromagnetic wave was detected when $\theta$ was 0° [Fig. 2(b)]. As a reference, we also measured the terahertz wave radiated from a 0.5- mm-thick (110)-oriented ZnTe crystal under the same experimental condition. The electric field of the terahertz wave of Co$_3$B$_7$O$_{13}$I with $\theta = 55°$ was evaluated to be about 1/300 of that of ZnTe.

To clarify the terahertz radiation mechanism, we measured the $\theta$ dependence of the electric field of the terahertz wave. To improve the signal-to-noise ratio, we measured the amplitude at 0 and -0.3 ps, and obtained the peak-to-peak amplitude ($E_{\text{THz}}^{\text{dif}}$). Figure 2(d) shows the $\theta$ dependence of $E_{\text{THz}}^{\text{dif}}$ at 0 and -0.3 ps. $E_{\text{THz}}^{\text{dif}}$ reaches the maximum at 55° and its sign is changed by rotating the sample by 90°. Furthermore, $E_{\text{THz}}^{\text{dif}}$ becomes zero at $\theta = 0, 90,$ and 180°. When optical rectification is the dominant process for the terahertz radiation, $E_{\text{THz}}^{\text{dif}}$ should be proportional to $P^{(2)}$ in Eq. (2), as discussed in Sect. 2. In the present experimental setup [Fig. 2(a)], we detected only the $X$-axis component of the terahertz wave. Thus, $E_{\text{THz}}^{\text{dif}}$ can be expressed as

$$E_{\text{THz}}^{\text{dif}} \propto -3\varepsilon_0 d_{14} E_0^2 \sin^2\theta \cos\theta. \qquad (6)$$

The solid line in Fig. 2(d) shows the fitting result using Eq. (6), which well reproduces the experimental result. Thus, we conclude that terahertz radiation in the non-centrosymmetric paraelectric phase of $Co_3B_7O_{13}I$ is due to the optical rectification.

### 4. 2. Terahertz radiation in non-centrosymmetric ferroelectric phase

Figure 3(a) shows the experimental setup for the non-centrosymmetric ferroelectric phase. The measurements were performed at 10 K. In this experiment, the $[001]_{cub}$ direction of the crystal was set parallel to the $X$-axis. We rotated the polarization of femtosecond laser pulses with a half-wave plate; $\phi = 0°$ corresponds to $E_0 \parallel [001]_{cub}$, whereas $\phi = 90°$ corresponds to $E_0 \parallel [1\bar{1}0]_{cub}$. According to the measurement of the polarized $\alpha$ spectra at 5 K, the penetration depth of femtosecond laser pulses was evaluated to be ~76 μm for $E_0 \parallel [001]_{cub}$ and ~71 μm for $E_0 \parallel [1\bar{1}0]_{cub}$. In the ferroelectric phase, we found terahertz radiation caused by the irradiation of femtosecond laser pulses. Figure 3(b) shows the electric field waveform of the radiated electromagnetic wave with $\phi = 0°$. The radiated wave mainly consists of a single-cycle terahertz pulse with a pulse width of ~2 ps. When $\phi$ was set to be 90° [Fig. 3(c)], the electric field of the terahertz wave was increased by a factor of 3, compared with that in $\phi = 0°$.

Next, we discuss the $\phi$ dependence of $E_{THz}^{dif}$ at 0 and -0.5 ps [Fig. 3(d)]. In contrast to the cubic phase [Fig. 2(d)], $E_{THz}^{dif}$ shows the maximum at 90°, with the minimum at 0°. In the orthorhombic phase, $E_{THz}^{dif}$ can be expressed as

$$E_{THz}^{dif} \propto \varepsilon_0 E_0^2 (d'_{32}\sin^2\phi + d'_{33}\cos^2\phi), \tag{7}$$

when the ferroelectric polarization is parallel to the $[001]_{cub}$ direction [see Eq. (4) in Sect. 2]. When the spontaneous polarization is parallel to $[100]_{cub}$ and $[010]_{cub}$ directions [see Eq. (5) in Sect. 2], $E_{THz}^{dif}$ is given as

$$E_{THz}^{dif} \propto \varepsilon_0 E_0^2 2\sqrt{2}(d''_{15} - d''_{24})\sin^2\phi + 2(d''_{15} + d''_{24})\sin 2\phi. \qquad (8)$$

As detailed in Sect. 4.3, ferroelectric domains oriented in the $[010]_{cub}$ and $[100]_{cub}$ directions simultaneously exist along the propagation direction of femtosecond laser pulses, i.e., $[\bar{1}\bar{1}0]_{cub}$ direction. Thus, $E_{THz}^{dif}$ follows the average of Eqs. (7) and (8). The solid line in Fig. 3(d) shows the fitting result, which well reproduces the measured $E_{THz}^{dif}$ with $d'_{33}/d'_{32} \sim 3$ ($d''_{15}$ and $d''_{24}$ are negligibly small at least by a factor of 100, compared with $d'_{33}$). Thus, optical rectification process is dominant for the terahertz radiation in the ferroelectric phase as in the paraelectric phase. We measured the temperature dependence of $E_{THz}^{dif}$ with $\phi = 0°$ ($d_{33}$) and confirmed that $d_{33}$ becomes zero in the paraelectric phase above 193 K. By comparing the maximum value of $E_{THz}^{dif}$ at 10 K with that at 300 K, $|d_{14}|$ was evaluated to be comparable to $|d'_{33}|$ with the opposite sign.

## 4. 3. Visualization of ferroelectric domains using emission of terahertz radiation

Here, we show the result of ferroelectric domain imaging by detecting the emitted terahertz wave. To detect ferroelectric polarization at each position as a vector, we performed vector analysis of radiated terahertz waves. In this experiment, $E_0$ of femtosecond laser pulses was set parallel to the $X$-axis, i.e., $E_0 \parallel [001]_{cub}$. In this configuration, the $[001]_{cub}$-oriented ferroelectric domain generates terahertz waves by the $d'_{33}$ component, whereas $[100]_{\text{cub}}$- and $[010]_{\text{cub}}$-oriented ferroelectric domains generate by the $d''_{31}$ and $d''_{32}$ components.

For the vector analysis, we set the angle of the wire grid polarizer to +45 or −45° with respect to the *X*-axis, which are illustrated in Fig. 4(a). In this setup, the *X* and *Y* components of the polarization can be expressed as

$$P_X \propto E_{+45°} + E_{-45°}, \qquad (9)$$

$$P_Y \propto E_{+45°} - E_{-45°}, \qquad (10)$$

respectively. $E_{+45°}$ and $E_{-45°}$ are the amplitudes obtained with +45 and −45°, respectively. Using the two images, the vector image of the electric polarization was constructed [box area in Fig. 4(a)]. Figure 4(b) shows the vector image in the same area of the visible image shown in Fig. 4(c). The direction of an arrow indicates the direction of the electric polarization at each position, while the length of an arrow indicates the magnitude of the electric polarization. In the lower region, $[1\bar{1}0]_{\text{cub}}$-oriented electric polarization was observed; this can be explained by the presence of $[100]_{\text{cub}}$- or $[0\bar{1}0]_{\text{cub}}$-oriented ferroelectric domains, as illustrated in Fig. 4(d). In the upper region with an area of ~500 × 500 μm², electric polarization is tilted in the (110) plane and not polarized along the direction of spontaneous polarization, i.e., $[100]_{\text{cub}}$, $[010]_{\text{cub}}$, and $[001]_{\text{cub}}$. Furthermore, note that electric polarizations gradually rotated counterclockwise from the lower to the upper regions. This type of ferroelectric domain, i.e., Neelwall in the case of magnetism, is not allowed, since the divergence of the polarization becomes nonzero in a large area of the crystal. Here, we discuss the observed vector orientations in terms of the presence of a ferroelectric 90° domain wall of the $(101)_{\text{cub}}$ plane. In previous birefringence measurements of $Ni_3B_7O_{13}Br$ and $Ni_3B_7O_{13}Cl$,[15] ferroelectric 90° domain walls were observed in $(011)_{cub}$, $(112)_{cub}$, $(\bar{1}11)_{cub}$, $(111)_{cub}$, and $(010)_{cub}$ planes. Thus, it is reasonable to consider the presence of a ferroelectric 90° domain wall of the $(011)_{cub}$ plane [or equivalently, $(101)_{cub}$ plane] in $Co_3B_7O_{13}I$ as well [Fig. 4(d)]. In this case, a ferroelectric 90° domain wall exists along the propagation direction of femtosecond laser pulses, i.e., $[\bar{1}\bar{1}0]_{cub}$ direction, and thus the detected terahertz waves become superposition of the generated terahertz waves in $[001]_{cub}$- and $[100]_{cub}$- (or $[0\bar{1}0]_{cub}$-) polarized ferroelectric domains, as schematically shown in Fig. 4(e). Consequently, the polarization of the detected terahertz wave becomes tilted in the $(110)_{cub}$ plane [Fig. 4(d)]; the tilting angle depends on the volume fraction of each ferroelectric domain. The tilting of the ferroelectric 90° domain wall also causes a smooth change in the volume fraction of ferroelectric domains along the $[1\bar{1}0]_{cub}$ direction [Fig. 4(d)], resulting in the gradual change in the vector orientations observed here.

## 5. Summary

Upon irradiation of femtosecond laser pulses, we observed terahertz radiation induced by optical rectification in the non-centrosymmetric paraelectric and ferroelectric phases of $Co_3B_7O_{13}I$. We visualized ferroelectric domains by mapping out the polarization vector of the radiated terahertz wave. In a large area of the crystal, we observed terahertz waves tilted in the (110) plane, which can be explained by the presence of a ferroelectric 90° domain wall of the $(101)_{cub}$ plane. Since the terahertz vector imaging method has an advantage of bulk- and phase-sensitive detection, it could be widely applied to ferroelectric domain imaging in various ferroelectrics.

## Acknowledgements

This work was partly supported by the Murata Foundation and Sumitomo Foundation, and by a Grant-in-Aid by Ministry of Education, Culture, Sports, Science and Technology (MEXT) (Nos. 25247049, 25247058, and 25-3372).

**Figure Captions**

Fig. 1. (Color online) Schematic illustrations of the crystal structure of $Co_3B_7O_{13}I$ in (a) cubic[17)] and (b) orthorhombic[18)] phases. (c) Solid and dotted circles indicate the positions of Co and I atoms in orthorhombic and cubic phases, respectively.

Fig. 2. (Color online) Terahertz radiation in the non-centrosymmetric paraelectric phase of $Co_3B_7O_{13}I$ by irradiation of femtosecond laser pulses, measured at room temperature. (a) Schematic illustration of the experimental setup. The electric field of the femtosecond laser pulses was set parallel to the *X*-axis in the laboratory coordinate. The angle $\theta$ was defined as the angle of the $[001]_{cub}$ direction of the crystal relative to the *X*-axis. Measured waveform

at (b) $\theta = 0°$ and (c) $\theta = 55°$. (d) Peak-to-peak amplitude at 0 and -0.3 ps as a function of $\theta$. The solid line shows the fitting result using Eq. (6).

Fig. 3. (Color online) Terahertz radiation in the non-centrosymmetric ferroelectric phase of $Co_3B_7O_{13}I$ by irradiation of femtosecond laser pulses, measured at 10 K. (a) Schematic illustration of the experimental setup. The $[001]_{cub}$ direction of the crystal was set parallel to the $X$-axis in the laboratory coordinate. The angle $\phi$ was defined as the angle of the electric field of the femtosecond laser pulses relative to the $X$-axis. Measured waveform at (b) $\phi = 0°$ and (c) $\phi = 90°$. (d) Peak-to-peak amplitude at 0 and -0.5 ps as a function of $\phi$. The solid line shows the fitting result using Eq. (7).

Fig. 4. (Color online) Schematic illustrations of (a) terahertz vector imaging using wire grid polarizer and vector basis (box area). (b) Terahertz vector image. (c) Optical image of the same area in (b). (d) Schematic illustrations of the possible ferroelectric domains and detected terahertz waves when we assume the presence of a ferroelectric 90° domain wall of the $(101)_{cub}$ plane. (e) Schematic illustration of the superposition of the generated terahertz waves emitted in each depth region.

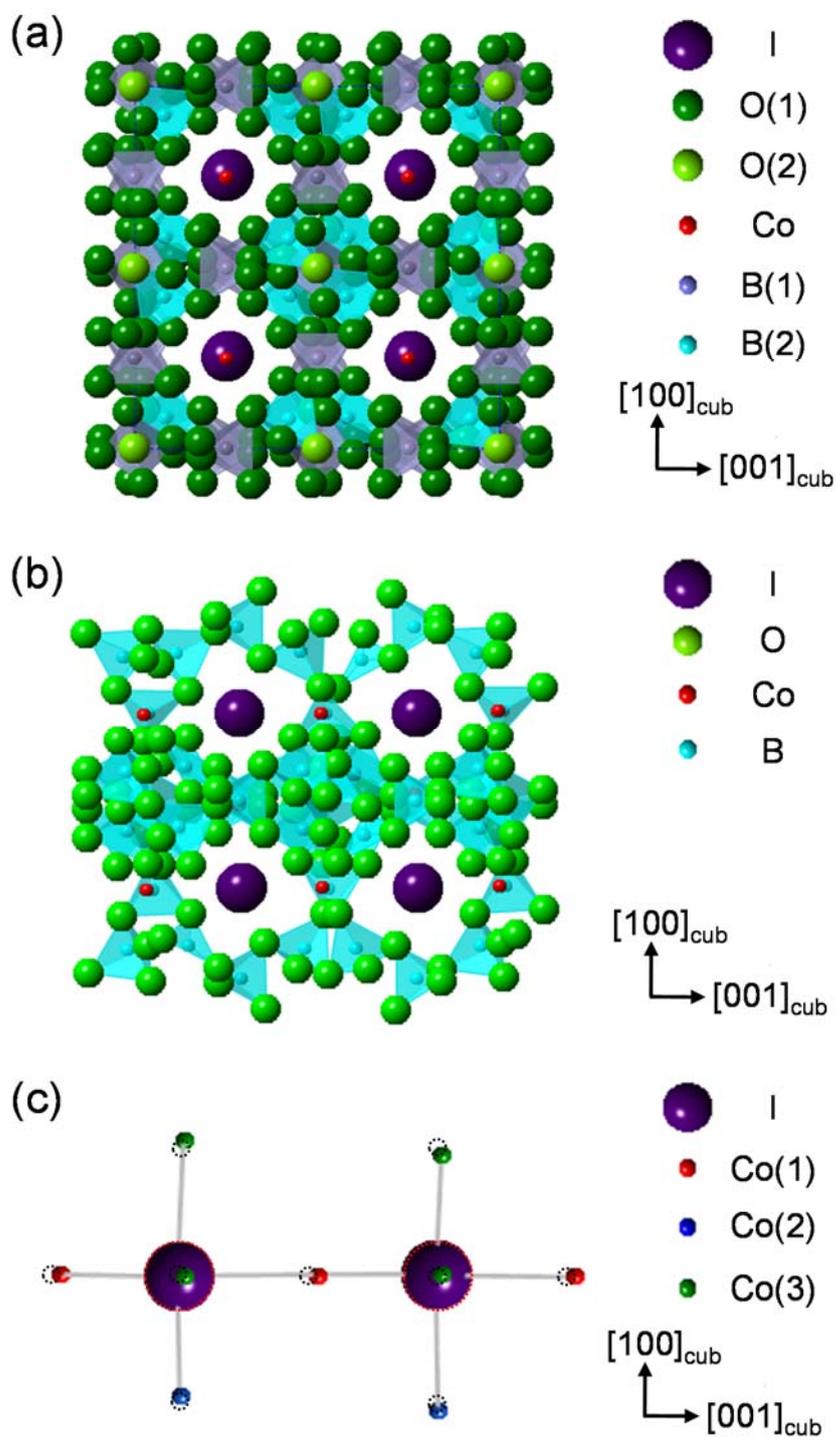

Fig. 1

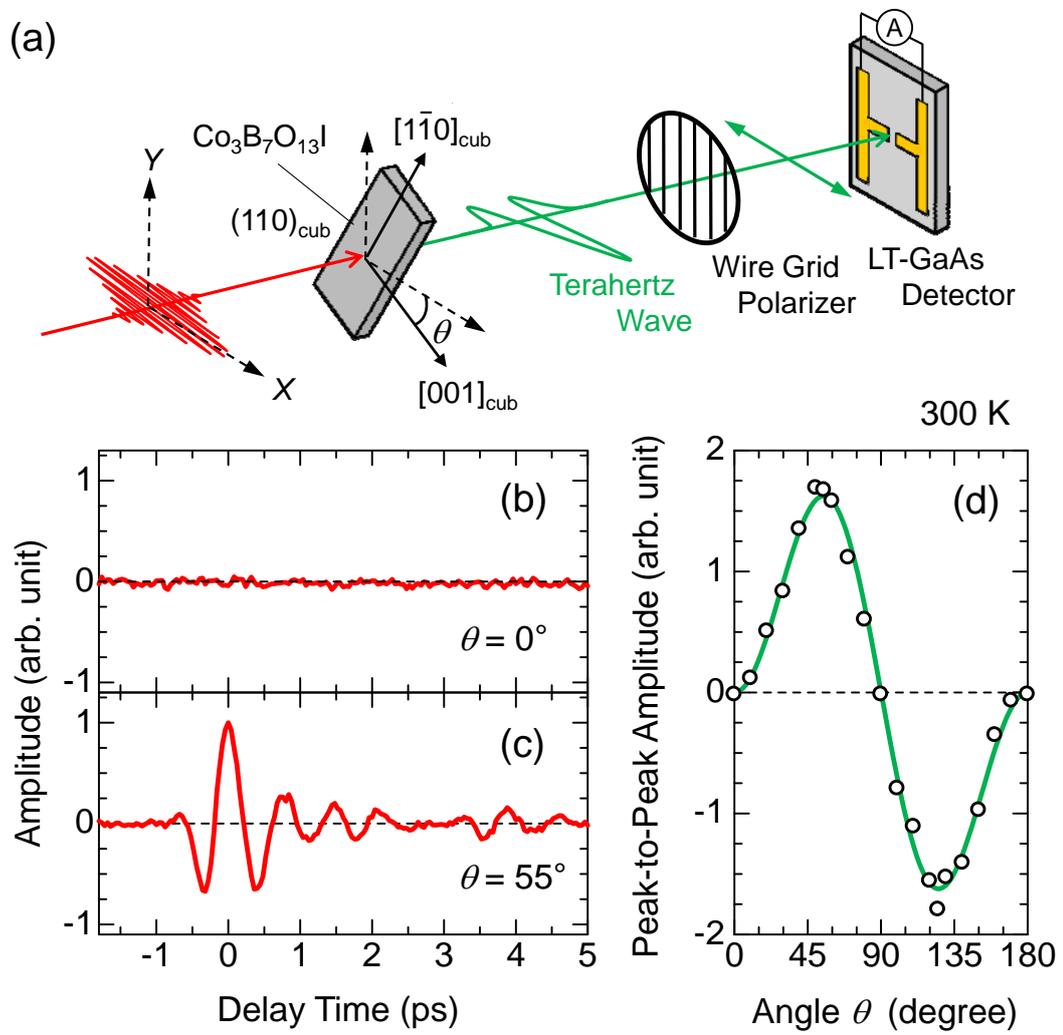

Fig. 2

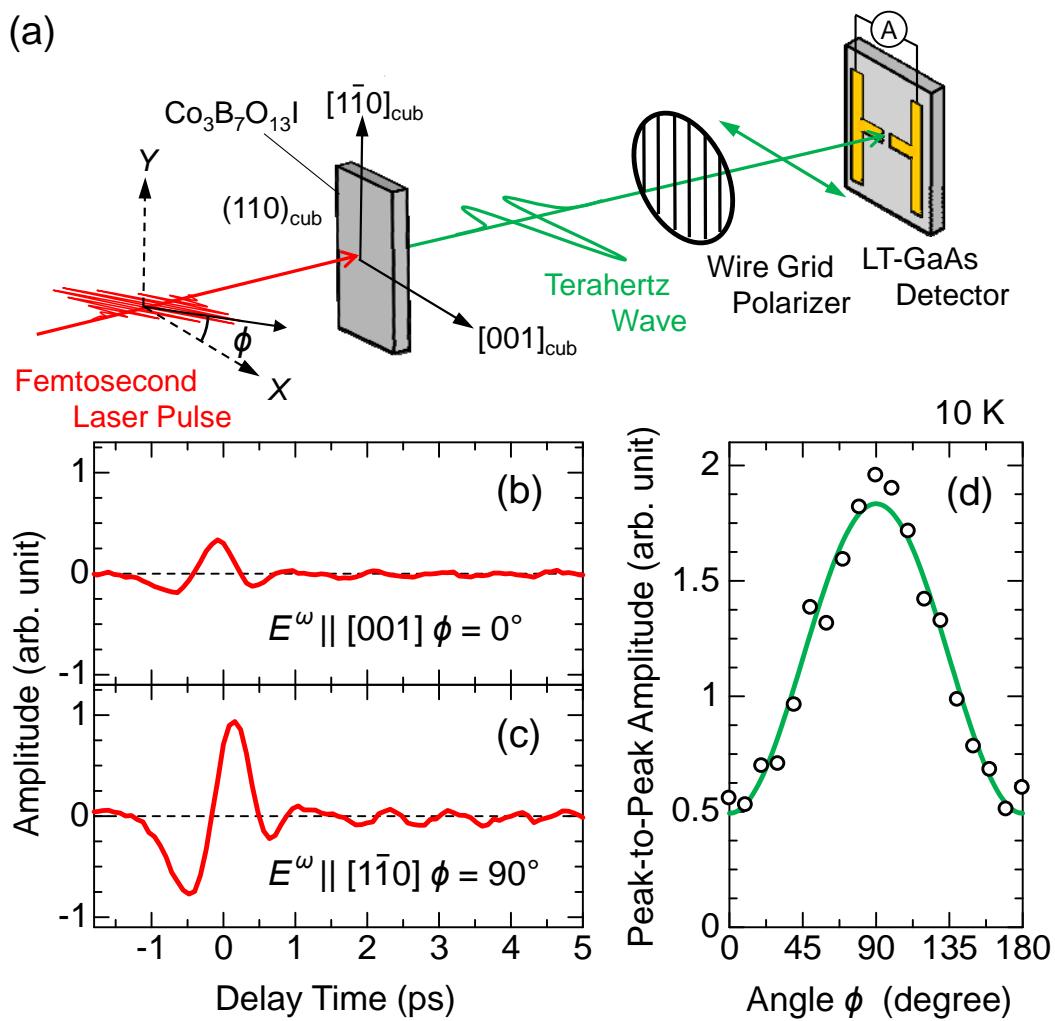

Fig. 3

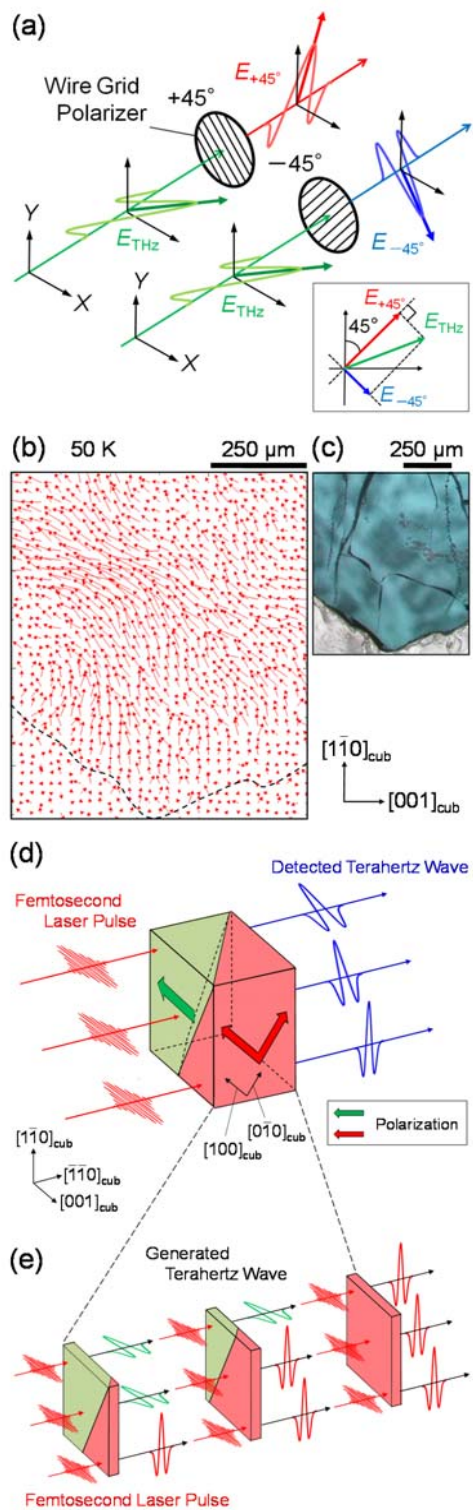

Fig. 4

19